\documentclass{jfm}
\usepackage{graphicx}
\usepackage{epstopdf, epsfig}
\usepackage[textwidth=0.55in]{todonotes}
\usepackage{amsmath}
\usepackage{amssymb}
\usepackage{amsfonts}
\usepackage{color}
\usepackage{graphicx}
\usepackage{bm}
\usepackage{epsfig}

\newcommand {\Wi}{W\!i}
\newcommand{\red}{\color{black}}

\title{Purely elastic linear instabilities in parallel shear flows with free-slip boundary conditions}

\author{Martin Lellep\aff{1}
  \corresp{\email{martin.lellep@ed.ac.uk}}, Moritz Linkmann\aff{2},
  Bruno Eckhardt\aff{3}\corresp{\red Deceased on the $7^\text{th}$ of August 2019. Professor Eckhardt took active part in conceiving this project and interpreting the free-slip pCF results. He has not seen the pPF part of this work that we hope we continued in line with his high scientific standards. All shortcomings should be attributed to A.M.}
\and Alexander Morozov\aff{1}}
\affiliation{\aff{1}SUPA, School of Physics and Astronomy, The University of Edinburgh, James Clerk Maxwell
Building, Peter Guthrie Tait Road, Edinburgh EH9 3FD, UK \aff{2}School of Mathematics and Maxwell Institute for Mathematical Sciences, University of Edinburgh, Edinburgh, EH9 3FD, United Kingdom \aff{3}Physics Department, Philipps-University of Marburg, D-35032 Marburg, Germany}

\begin{document}

\maketitle

\begin{abstract}
We perform a linear stability analysis of viscoelastic plane Couette and plane Poiseuille flows with free-slip boundary conditions. The fluid is described by the Oldroyd-B constitutive model, and the flows are driven by a suitable body force. We find that both types of flow become linearly unstable, and we characterise the spatial structure of the unstable modes. By performing a boundary condition homotopy from the free-slip to no-slip boundaries, we demonstrate that the unstable modes are directly related to the least stable modes of the no-slip problem, destabilised under the free-slip situation. We discuss how our observations can be used to study recently discovered purely elastic turbulence in parallel shear flows. 
\end{abstract}


\section{Introduction}

Stability of parallel shear flows of dilute polymer solutions is an outstanding problem of mechanics of complex fluids. While linear stability analyses performed with model viscoelastic fluids suggest that these flows are linearly stable \citep{Gorodtsov1967,Wilson1999}, it has been proposed that they can lose their stability through a finite-amplitude, sub-critical bifurcation \citep{Morozov2005,Morozov2007,Morozov2019}, similar to their Newtonian counterparts \citep{Grossmann2000}. Recent experiments performed in pressure-driven straight channels support this scenario: For sufficiently high flow rates, there exists a critical strength of flow perturbations required to destabilise the flow \citep{Pan2013}, thus demonstrating the sub-critical nature of the transition. Beyond the transition, the flow exhibits features of purely elastic turbulence \citep{Bonn2011,Pan2013,Qin2017}, previously only reported in geometries with curved streamlines \citep{Groisman2000,Steinberg2021}.

Despite recent progress, the exact origin of purely elastic turbulence remains unknown. One of the major obstacles in understanding its mechanism is the absence of direct numerical simulations of confined flows at sufficiently high flow rates due to the high Weissenberg number problem \citep{Owens2002}, which is of purely numerical origin. Although significant progress has been made in developing numerical methods capable of controlling this problem \citep{Alves2021}, direct numerical simulations relevant to experiments in parallel shear flows \citep{Bonn2011,Pan2013,Qin2017} are still lacking. Here, we propose to circumvent the high Weissenberg number problem by studying parallel shear flows with free slip boundary conditions that remove steep near-wall velocity gradients and significantly improve numerical stability.

The influence of wall slip on hydrodynamic stability of various flows has been extensively studied. The motivation behind these studies can be separated into two broad classes. The studies from the first class seek to approximate microscopically sound boundary conditions and thus address the question of linear stability of experimentally realisable flows. Such an approach has indicated, for instance, that the classical two-dimensional Tollmien–Schlichting transition in plane channel flow of Newtonian fluids is significantly suppressed by the presence of wall slip \citep{Gersting1974,Lauga2005,Min2005}, although the non-normal energy growth mechanism was shown to be less affected \citep{Lauga2005}. When adopted to shear flows of polymeric fluids, this approach revealed the existence of a short wave-length instability \citep{Black1996,Black1999,Black2001}, absent in the corresponding no-slip flows.

The second class of studies employs slip boundary conditions exclusively as the means to simplify the underlying no-slip boundary condition problem. Free-slip boundary conditions usually adopted in those works are rather artificial, yet they yield significantly more tractable mathematical problems. This approach was pioneered by \cite{Rayleigh1916} in the context of Rayleigh–B\'{e}nard instability. More recently, \cite{Waleffe1997} used free-slip boundary conditions to construct exact coherent states in parallel shear flows of Newtonian fluids that revolutionised our understanding of the transition to turbulence in those flows \citep{Eckhardt2007,Tuckerman2020,Graham2021}. The most important outcome of these studies is the observation that free-slip systems preserve the main phenomenology of their no-slip counterparts, albeit at increased values of the control parameter. Our work draws its motivation from this class of studies.

Here, we perform a linear stability analysis of plane Couette flow (pCF) and pressure-driven plane Poiseuille flow (pPF) with free-slip boundary conditions. 
We employ the Oldroyd-B model \citep{Bird1987} in the absence of inertia, and demonstrate numerically that these flows develop linear instabilities absent in their no-slip counterparts. Importantly, we observe that the main features of no-slip flows are preserved under the free-slip conditions, and we discuss how these instabilities can be employed to gain insight into the nature of elastic turbulence in parallel geometries.

\section{\label{sec:problem_setting}Problem setup}

We study incompressible flows of a model viscoelastic fluid confined between two infinite parallel plates. We introduce a Cartesian coordinate system $\{x_1,x_2,x_3\}$ oriented along the streamwise, gradient, and spanwise directions, respectively; the plates are located at $x_2=\pm L$. The dynamics of the fluid are assumed to be described by the Oldroyd-B model \citep{Bird1987},

\begin{subequations}
\begin{align}
%% Oldroyd B
&{\bm\tau} + \Wi \left[ \frac{\partial {\bm\tau}}{\partial t} + {\bm v}\cdot\nabla{\bm\tau} - \left(\nabla {\bm v}\right)^T\cdot{\bm\tau} - {\bm\tau}\cdot\left(\nabla {\bm v}\right)\right] = \left(\nabla {\bm v}\right) + \left(\nabla {\bm v}\right)^T, 
\label{eqn:oldroyd_b.constitutive} \\
%% Navier Stokes
&\qquad Re \left[ \frac{\partial {\bm v}}{\partial t} + {\bm v}\cdot\nabla{\bm v}  \right] = -\nabla p + \beta \nabla^2{\bm v} + (1-\beta)\nabla\cdot{\bm \tau}  + {\bm f}, 
\label{eqn:oldroyd_b.momentum} \\
%% Incompressibility
&\qquad\qquad\qquad\qquad  \qquad\qquad  \nabla\cdot {\bm v} = 0, 
\label{eqn:oldroyd_b.incompressibility}
\end{align}
\label{eqn:oldroyd_b}
\end{subequations} 
\kern-0.3em where $\bm v$, $\bm \tau$, and $p$ are the fluid velocity, polymeric Cauchy stress tensor, and the pressure, respectively, while $\left(\nabla {\bm v}\right)^T$ denotes the transpose of the velocity gradient tensor; $\bm f$ is a body force to be specified below. These equations are rendered dimensionless using the channel's half-width $L$ as the unit of length, the maximum velocity of the laminar profile $V_0$ as the unit velocity, and $L/V_0$ as the unit of time. The governing equations \eqref{eqn:oldroyd_b} feature three dimensionless groups, the Weissenberg number $\Wi=\lambda V_0/L$, Reynolds number $Re= \rho L V_0/\eta$, and the viscosity ratio $\beta=\eta_s/\eta$, where $\lambda$ is the Maxwell relaxation time of the fluid, $\rho$ is its density, while $\eta_s$ and $\eta$ are the solvent viscosity and the total viscosity of the fluid, respectively. Unless stated otherwise, below we focus on the purely elastic limit with $Re=0$.

The fluid is assumed to satisfy the following free-slip boundary conditions at the walls
\begin{align}
\partial_2 v_1 = v_2 = \partial_2 v_3 = 0, \qquad \text{for }x_2=\pm 1,
\label{eqn:1.boundary_conditions}
\end{align}
where $\partial_2$ denotes the spatial derivative in the gradient direction. In this work, we study unidirectional laminar profiles ${\bm v} = \{U(x_2),0,0\}$, where $U$ is chosen to resemble the shape and symmetry of either plane Poiseuille flow (pPF) or plane Couette flow (pCF), while satisfying the boundary conditions, Eq.\eqref{eqn:1.boundary_conditions},
\begin{align}
U(x_2) =
\begin{cases}
\sin(\pi x_2/2),   & \qquad\text{(pCF),} \\
\cos(\pi x_2/2)^2, & \qquad\text{(pPF).}
\end{cases}
\label{eq:laminar_profile}
\end{align}
The corresponding laminar components of the polymeric stress tensor are given by $\tau_{11} = 2 \Wi\left(\partial_2 U(x_2)\right)^2$ and $\tau_{12} = \partial_2 U(x_2)$, with all other components being zero. To drive such flows in the presence of free-slip boundary conditions, we add the body force ${\bm f} = \{-\partial_2^2 U(x_2),0,0\}$ to the r.h.s. of Eq.\eqref{eqn:oldroyd_b.momentum}.

To perform a linear stability analysis of these flows, we linearise Eqs.\eqref{eqn:oldroyd_b} around the laminar profile given above. In view of the viscoelastic analogue of the Squire's theorem \citep{Bistagnino2007}, that also holds for the free-slip boundary conditions studied here, we consider two-dimensional modal perturbations $\{\delta {\bm v}(x_2), \delta {\bm \tau}(x_2), \delta p(x_2) \} e^{i k_1 x_1} e^{\sigma t}$ of the velocity, stress, and pressure, respectively. Here, $k_1$ sets the spatial scale of the perturbation, and $\sigma=\sigma_r + i \sigma_i$ denotes a complex temporal eigenvalue, with $\sigma_r$ and $\sigma_i$ being its real and imaginary part, respectively. The $x_2$ dependence of the perturbations are expanded in Chebyshev polynomials, and the linearised equations of motion are solved with a fully spectral Chebyshev-Tau method \citep{canutobook} by employing the generalised eigenvalue solver {\tt eig} from the scientific library SciPy for Python \citep{2020SciPy-NMeth}.

\section{\label{sec:results}Results}

\begin{figure}
	\centering
	\includegraphics[scale=0.7]{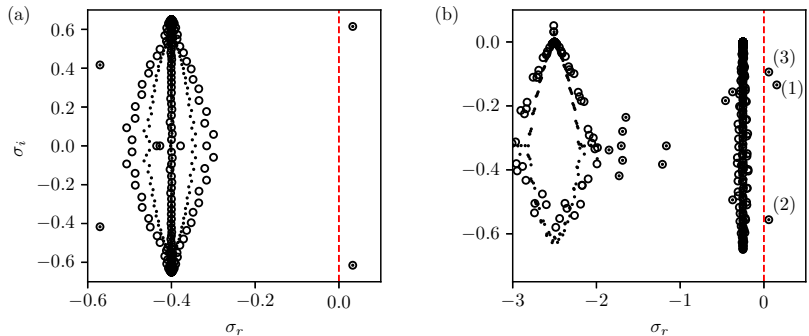}
	\caption{
		Eigenvalue spectra for $\beta=0.1$ and $k_1=0.65$ for (a) pCF at $Wi=2.5$, and (b) pPF at $Wi=4.0$. The dashed red lines denote marginal stability. {\red Both figures compare spectra calculated with $50$ (open circles) and $100$ (dots) Chebyshev polynomials.}}
	\label{fig:spectrum}
\end{figure}

The no-slip pCF and pPF of Oldroyd-B fluids are linearly stable for a wide range of $\beta$ and $\Wi$ \citep{Gorodtsov1967,Wilson1999}. Recently \cite{Khalid2021} reported a purely elastic linear instability in no-slip pPF, however, its region of existence is confined to a small and, most likely, experimentally inaccessible part of parameter space with $\beta>0.9$ and $\Wi>O(10^3)$. Surprisingly, here we find that free-slip pCF and pPF are linearly unstable for a wide range of $\beta$ and $\Wi$.

In Fig.\ref{fig:spectrum}, we present examples of such instabilities by plotting the eigenvalue spectra for pCF and pPF. Similar to the no-slip problem \citep{Wilson1999}, the spectra consist of discrete physical eigenvalues and two lines of eigenvalues corresponding to continuous spectra with $\sigma_r=-1/\Wi$ and $\sigma_r = -1/(\beta \Wi)$; the finite-resolution approximations to the latter take the balloon-like shapes visible in Fig.~\ref{fig:spectrum}. The most profound feature of both spectra in Fig.~\ref{fig:spectrum} is that all leading eigenvalues have positive real parts, indicating the presence of a linear instability. 
%In striking contrast to their no-slip counterparts, two of the three leading eigenvalues in pPF have the same real part for all values of the parameters we considered.
{\red Surprisingly, two of the three leading eigenvalues in pPF have numerically almost identical real parts for all values of the parameters we considered (see also Fig.\ref{fig:eigenvalue_homotopy} below).}

To demonstrate numerical convergence, in Fig.\ref{fig:spectrum}, we present the eigenvalue spectra for two spectral resolutions, with $M=50$ and $M=100$ Chebyshev polynomials, respectively, and observe that the physical eigenvalues are well-converged. All calculations presented below are therefore carried out with $M=50$. While this spectral resolution is quite low to yield converged results in the no-slip setting \citep{Wilson1999,Khalid2021}, it is clearly sufficient for the free-slip boundary conditions. This observation supports our original rationale that the free-slip boundary conditions simplify numerical simulations of such flows by removing steep near-wall velocity gradients. 

{\red
\begin{figure}
	\centering
	\includegraphics[scale=0.7]{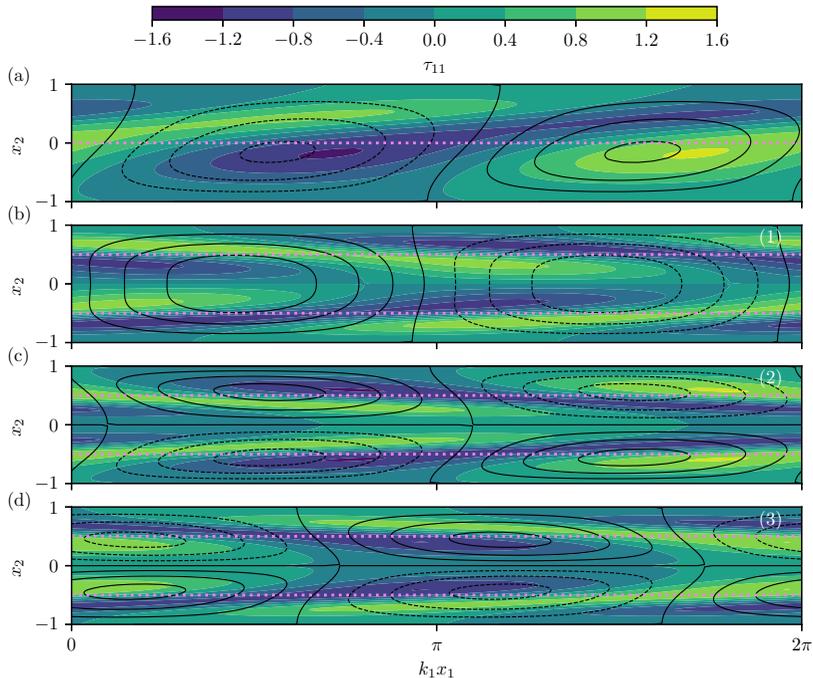}
	\caption{Spatial profiles of the polymeric stress $\delta\tau_{11}$ (colour) and the streamfunction (contours with negative values denoted by dashed lines) corresponding to the unstable eigenvalues in Fig.\ref{fig:spectrum} for $\beta=0.1$ and $k_1=0.65$. (a) pCF at $\Wi=2.5$. (b)-(d) pPF eigenmodes (1)-(3) at $\Wi=4.0$, see Fig.\ref{fig:spectrum}(b). {\red The violet dotted lines denote the positions of the maxima of the base components $\vert \tau_{12}\vert$ and $\vert \tau_{11}\vert$.}	}
	\label{fig:profiles}
\end{figure}
}

For no-slip boundary conditions, the least stable eigenmode of pCF and one of the three leading eigenmodes of pPF are mostly localised in the vicinity of the confining boundaries \citep{Gorodtsov1967,Wilson1999}, and are thus referred to as \emph{wall modes}, while the other two pPF modes are mostly localised in the bulk \citep{Wilson1999,Khalid2021}, i.e. they are \emph{centre modes}. In Fig.\ref{fig:profiles} we present the spatial profiles corresponding to the unstable free-slip eigenvalues discussed above, and observe that the wall {\it versus} centre mode distinction is significantly weakened. Although the symmetries of the stress distribution and the streamfunction are still different between these modes (compare the vortex structure in Fig.\ref{fig:profiles}(a)-(b) with (c)-(d), for instance), all modes now have significant presence throughout the domain. Notably, the fact that all these modes can be simultaneously unstable indicates the presence of several, rather different instability mechanisms.

\begin{figure}
	\centering
	\includegraphics[scale=0.7]{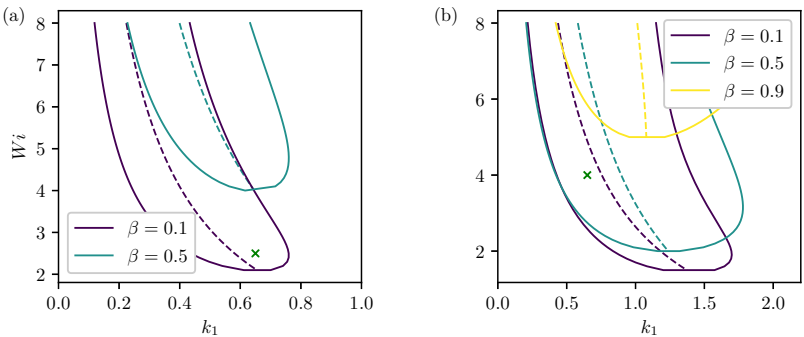}
	\caption{
		Neutral stability curves (solid lines) and the position of the most unstable eigenvalue maxima (dashed lines) for (a) pCF and (b) pPF. The green crosses denote the parameters used in Fig.~\ref{fig:spectrum}.
	}
	\label{fig:neutral_stability_curves}
\end{figure}

\begin{figure}
	\centering
	\includegraphics[width=0.7\columnwidth]{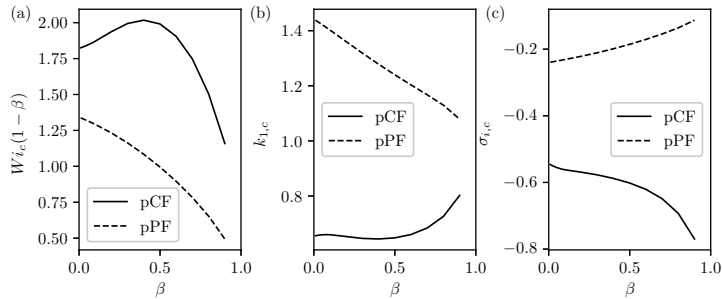}
	\caption{
		(a) The critical Weissenberg number $Wi_c$ as a function of $\beta$. {\red The corresponding critical wave number $k_{1,c}$ (b)  and  imaginary part of the critical eigenvalue $\sigma_{i,c}$ as a function of $\beta$ (c).}
	}
	\label{fig:criticalWiK1_beta}
\end{figure}

In Fig.\ref{fig:neutral_stability_curves}  we present the neutral stability curves for various values of $\beta$, while Fig.\ref{fig:criticalWiK1_beta} shows the critical Weissenberg number $\Wi_{c}$ and the corresponding critical wave number $k_{1,c}$ as functions of $\beta$. The neutral stability curves are skewed towards small values of  $k_1$ in a way reminiscent of the  elasto-inertial \citep{Chaudhary2019,Khalid2021a} and purely elastic modes \citep{Khalid2021} discovered recently. Despite these similarities, we were unable to use the re-scaling procedure proposed in those works to collapse all our neutral stability data on a single master curve. Further, in Fig.\ref{fig:criticalWiK1_beta}(a) we observe that $\Wi_c$ increases rapidly as $\beta\to 1$, indicating that a sufficiently strong velocity-stress coupling is required to drive the instability. We also note that $\Wi_c$ is significantly smaller for pPF than for pCF, and that, surprisingly, the critical wave number $k_{1,c}$ depends in different ways on $\beta$ for these flows. As can be seen in Fig.\ref{fig:criticalWiK1_beta}(b), in all cases, $k_{1,c}\sim O(1)$, indicating an instability lengthscale comparable with the wall-to-wall distance.

\begin{figure}
	\centering
	\includegraphics[scale=0.7]{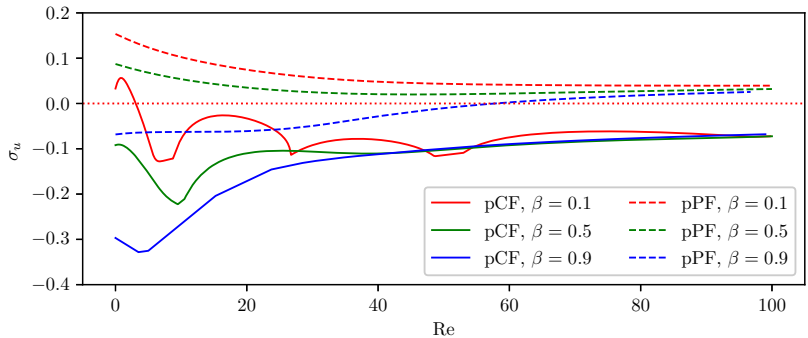}
	\caption{
		Influence of weak inertia on linear instability of pCF and pPF at $Wi=2.5$ and $Wi=4.0$, respectively, at $k_1=0.65$ as denoted by the green cross in Fig.~\ref{fig:neutral_stability_curves}. The dotted red line denotes marginal stability.
	}
	\label{fig:weak_inertia}
\end{figure}

To assess the influence of weak inertia, in Fig.\ref{fig:weak_inertia} we plot the leading eigenvalue of pCF and pPF as a function of the Reynolds number. We see that while small amounts of inertia have stabilising effect on pCF, the low-$\beta$ instabilities in pPF are only mildly suppressed. For pPF at sufficiently high $\beta$, the trend reverses, and we observe a stable purely elastic eigenmode becoming unstable for $Re>0$. While a full $\Wi-\beta-Re$ critical surface is required to draw a definite conclusion regarding the influence of inertia on each mode, we note here that our observations are in line with the recent work on elasto-inertial instabilities \citep{Chaudhary2019,Khalid2021a}.

%The spectra presented in Fig.\ref{fig:spectrum} bear a striking resemblance to the spectra of the corresponding flows of Oldroyd-B fluids with no-slip boundary conditions \citep{Gorodtsov1967, Wilson1999}. The main difference between the two types of boundary conditions is the position of the leading eigenvalues that are stable, in the no-slip case, but move towards the positive values of $\sigma_r$, in the free-slip case. To firmly establish this connection, we perform a  homotopy between the two types of boundary conditions. To this end, we introduce the following laminar profile that depends on a parameter $\alpha$, 

The spectra presented in Fig.\ref{fig:spectrum} bear a striking resemblance to the spectra of the corresponding flows of Oldroyd-B fluids with no-slip boundary conditions \citep{Gorodtsov1967, Wilson1999} but for the position of the leading eigenvalues that appear to move to positive values of $\sigma_r$ in the free-slip case. To establish a connection between these eigenvalues and their no-slip counterparts, we perform a homotopy between the two types of boundary conditions. To this end, we introduce the following laminar profile that depends on a parameter $\alpha$, 
\begin{align}
U(x_2) =
\begin{cases}
\alpha \sin(\pi x_2/2) + (1-\alpha) x_2 ,   & \qquad\text{(pCF),} \\
\alpha \cos(\pi x_2/2)^2 + (1-\alpha) (1-x_2^2) , & \qquad\text{(pPF),}
\end{cases}
\label{eq:homotopy.laminar_profile}
\end{align}
that interpolates between Eq.\eqref{eq:laminar_profile} ($\alpha=1$) and the classical no-slip pCF and pPF profiles ($\alpha=0$). A similar interpolation is imposed on the boundary conditions
\begin{align}
\alpha \partial_2 \delta v_1 \pm (1-\alpha) \delta v_1 = v_2 = \alpha \partial_2 \delta v_3 \pm (1-\alpha) \delta v_3 = 0, \qquad \text{for }x_2=\pm 1.
\label{eqn:1.boundary_conditions.homotopy}
\end{align}
In Fig.~\ref{fig:eigenvalue_homotopy} we show the superimposed spectra for this laminar profile as a function of $\alpha$ for pCF and pPF. The results of the homotopy procedure clearly indicate that the unstable modes do not appear either from infinity or the continuous spectra, but instead are continuously connected to the well-known least stable eigenvalues of no-slip parallel shear flows \citep{Gorodtsov1967, Wilson1999}.

\begin{figure}
	\centering
	\includegraphics[scale=0.7]{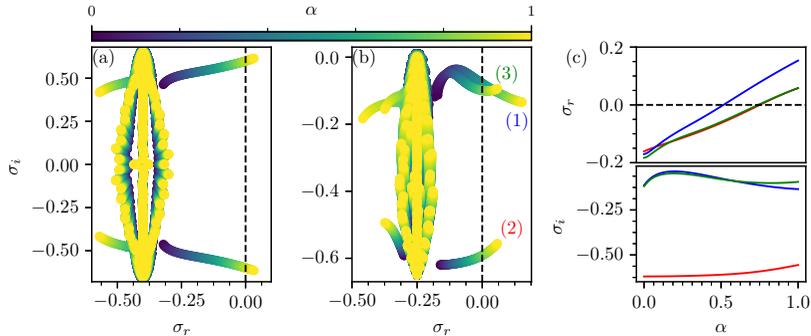}
	\caption{
		{\red
		(a,b) Eigenvalue spectrum of system with continuously varied boundary conditions from free-slip (yellow) to no-slip (violet) for (a) pCF at $Wi=2.5$ and (b) pPF at $Wi=4.0$. Data shown in both panels were calculated for $\beta=0.1$, $Re=0.0$ and $k_1=0.65$. The black dashed line denotes marginal stability. Each spectrum is colour-coded according to the value of the homotopy parameter $\alpha$. (c) Real and imaginary parts of the three leading eigenvalues from (b) as a function of $\alpha$. The dashed line denotes the instability threshold.	 The leading pCF eigenvalues (not shown) become stable for $\alpha<0.88$, while the leading pPF eigenvalues (1), (2), and (3) become stable for $\alpha<0.52$, $0.75$, and $0.74$, respectively.}
	}
	\label{fig:eigenvalue_homotopy}
\end{figure}

\section{\label{sec:conclusions}Conclusions}

In this work, we demonstrate that free-slip pCF and pPF of Oldroyd-B fluids are linearly unstable. The instabilities we observe are caused by neither the curvature of the base flow streamlines, where hoop stresses are known to lead to purely elastic instabilities \citep{McKinley1996}, nor by steep gradients of the material properties, as in the co-extrusion problems \citep{Hinch1992}. Suggestively, the base velocity profiles considered here change the sign of their second derivatives inside the domain, which might indicate a connection to Rayleigh's inflection point instability criterion for inviscid fluids \citep{Drazin2004}. A similar situation occurs in purely elastic Kolmogorov flows \citep{boffetta2005}, and it would be interesting to examine whether the instabilities observed there are related to those found in this work. By performing the homotopy between free-slip and no-slip boundary conditions, we demonstrate that the unstable eigenmodes are directly related to the least stable modes of the no-slip system. The exact way by which the introduction of an inflection point into the base profile would lead to destabilisation of the no-slip modes is currently unknown.

While the body forcing and the boundary conditions considered here are probably difficult, if not impossible, to realise in practice, the main importance of our work is not related to its direct experimental relevance. Instead, our contribution here is related to the observation that the free-slip problem considered here is significantly simpler to address numerically than its no-slip counterpart. Even at the linear level, a significantly lower number of Chebyshev modes is needed to resolve the eigenspectrum than in the no-slip case. A similar simplification is expected to also occur in fully non-linear direct numerical simulations and to alleviate the High Weissenberg number problem plaguing such studies. By rendering the leading eigenmodes unstable, free-slip flows offer an easier route to study strongly sub-critical structures that propagate from those eigenmodes in the no-slip case \citep{Morozov2005,Morozov2007,Morozov2019}. 

Once coherent states are resolved in direct numerical simulations of free-slip flows, a boundary condition homotopy, similar to the one employed here, can be used to track those solutions back to the no-slip conditions \citep{Waleffe2003}. Recent work on elasto-inertial parallel shear flows \citep{Garg2018,Chaudhary2019,Khalid2021a,Chaudhary2021} 
associates linear instabilities at high $\Wi$ and $Re$ with the very leading eigenvalues studied here.
As already mentioned above, the same instability can be traced to the purely elastic case, $Re=0$, for sufficiently large values of $\beta$ and $\Wi$ \citep{Khalid2021}. It has also been shown that these instabilities are sub-critical in nature \citep{Page2020}. Taken together with the original proposal of  \cite{Morozov2005,Morozov2007,Morozov2019}, and the experimental and numerical studies of non-linear states in elasto-inertial flows \citep{Samanta2013,Dubief2013,Sid2018,Choueiri2018,Lopez2019}, this points towards the existence of exact coherent states related to the least stable eigenvalues, even though they are linearly stable for most values of $\beta$ and $\Wi$. Our work suggests using a homotopy from free-slip flows to discover these structures.

\begin{figure}
	\centering
	\includegraphics[scale=0.7]{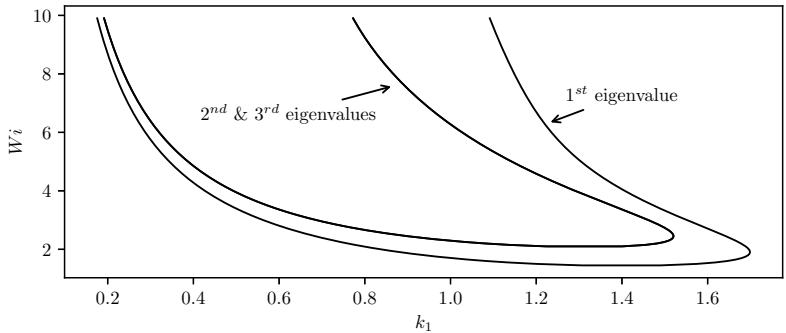}
	\caption{
		Neutral stability curves of the first three eigenvalues of pPF for $\beta=0.1$. As mentioned in Section \ref{sec:results}, the eigenvalues 2 and 3 have the same real parts, and their instability regions coincide.
		}
	\label{fig:neutral_curve_three_eigs}
\end{figure}

Finally, we note that our results suggest that all three leading eigenvalues of free-slip pPF can become unstable, see Fig.\ref{fig:spectrum}, for example. In elasto-inertial flows, there is an ongoing discussion whether the most dynamically relevant coherent structures are related to wall or centre-line modes \citep{Shekar2019,Datta2021preprint}. In Fig.\ref{fig:neutral_curve_three_eigs}, we present the neutral stability curve for the first three eigenvalues and observe that by selecting the wavenumber and the Weissenberg number, one can control the type of the instability thus produced. As Fig.\ref{fig:profiles} suggests, the different spatial symmetries of those modes should result in structures that connect to either wall or centre-line modes, when traced back to the no-slip conditions, thus allowing one to assess their relative importance independently. 

\vspace{0.5cm}

\textbf{Declaration of Interests.} The authors report no conflict of interest.

\bibliographystyle{jfm}
\bibliography{my_library}

\end{document}